\documentclass[numberedappendix, appendixfloats]{emulateapj}
\usepackage{graphicx}
\usepackage[space]{grffile}
\usepackage{latexsym}
\usepackage{amsfonts,amsmath,amssymb}
\usepackage{url}
\usepackage[utf8]{inputenc}
\usepackage{hyperref}
\hypersetup{colorlinks=false,pdfborder={0 0 0}}
\usepackage{textcomp}
\usepackage{longtable}
\usepackage{multirow,booktabs}
\usepackage[english]{babel}
\usepackage{natbib}
\usepackage{url}

\newcommand{\NS}{\mbox {\it NuSTAR}}

\usepackage{natbib}

\shorttitle{NuSTAR Solar Observations}
\shortauthors{Brian Grefenstette}

\begin{document}


\title{The First Focused Hard X-ray Images of the Sun with NuSTAR}


\author{
Brian~W.~Grefenstette\altaffilmark{1},
Lindsay~Glesener\altaffilmark{2,3},
S{\"a}m~Krucker\altaffilmark{3,4},
Hugh~Hudson\altaffilmark{3,5},
Iain~G.~Hannah\altaffilmark{5},
David~M.~Smith\altaffilmark{6},
Julia~K.~Vogel\altaffilmark{7},
Stephen~M.~White\altaffilmark{8},
Kristin~K.~Madsen\altaffilmark{1},
Andrew~J.~Marsh\altaffilmark{6},
Amir~Caspi\altaffilmark{9},
Bin~Chen\altaffilmark{10},
Albert~Shih~\altaffilmark{11},
Matej~Kuhar~\altaffilmark{4},
Steven~E.~Boggs\altaffilmark{3},
Finn E. Christensen\altaffilmark{12},
William W. Craig\altaffilmark{3, 13},
Karl~Forster\altaffilmark{1},
Charles J. Hailey\altaffilmark{14},
Fiona~A.~Harrison\altaffilmark{1},
Hiromasa~Miyasaka\altaffilmark{1},
Daniel Stern\altaffilmark{15},
William W. Zhang\altaffilmark{16}
}

\def \cahill {Cahill Center for Astrophysics, 1216 E. California Blvd, California Institute of Technology, Pasadena, CA 91125, USA}
\altaffiltext{1}{\cahill}
\altaffiltext{2}{School of Physics and Astronomy, University of Minnesota - Twin Cities , Minneapolis, MN 55455, USA}
\altaffiltext{3}{Space Sciences Laboratory, University of California, Berkeley, CA 94720, USA}
\altaffiltext{4}{University of Applied Sciences and Arts Northwestern Switzerland, CH-5210 Windisch, Switzerland}
\altaffiltext{5}{SUPA School of Physics and Astronomy, University of Glasgow, Glasgow G12 8QQ, UK}
\altaffiltext{6}{Physics Department and Santa Cruz Institute for Particle Physics, University of California, Santa Cruz, 1156 High Street, Santa Cruz, CA 95064, USA}
\altaffiltext{7}{Physics Division, Physical and Life Sciences Directorate, Lawrence Livermore National Laboratory, Livermore, CA 94550, USA}
\altaffiltext{8}{Air Force Research Laboratory, Albuquerque, NM, USA}
\altaffiltext{9}{Southwest Research Institute, Boulder, CO 80302, USA}
\altaffiltext{10}{Harvard-Smithsonian Center for Astrophysics, Cambridge, MA 02138, USA}

\altaffiltext{11}{Solar Physics Laboratory, NASA Goddard Space Flight Center, Greenbelt, MD 20771, USA}

\altaffiltext{12}{DTU Space, National Space Institute, Technical University of Denmark, Elektrovej 327, DK-2800 Lyngby, Denmark}
\altaffiltext{13}{Lawrence Livermore National Laboratory, Livermore, CA 94550, USA}
\altaffiltext{14}{Columbia Astrophysics Laboratory, Columbia University, New York, NY 10027, USA }
\altaffiltext{15}{Jet Propulsion Laboratory, California Institute of Technology, Pasadena, CA 91109, USA}
\altaffiltext{16}{Astrophysics Science Division, NASA/Goddard Space Flight Center, Greenbelt, MD 20771, USA}

\email{bwgref@srl.caltech.edu}




\bibliographystyle{apj}

\begin{abstract}

We present results from the the first campaign of dedicated solar observations undertaken by the \textit{Nuclear Spectroscopic Telescope ARray} ({\em NuSTAR}) hard X-ray telescope. Designed as an astrophysics mission, {\em NuSTAR} nonetheless has the capability of directly imaging the Sun at hard X-ray energies ($>$3~keV) with an increase in sensitivity of at least two magnitude compared to current non-focusing telescopes. In this paper we describe the scientific areas where \textit{NuSTAR} will make major improvements on existing solar measurements. We report on the techniques used to observe the Sun with \textit{NuSTAR}, their limitations and complications, and the procedures developed to optimize solar data quality derived from our experience with the initial solar observations. These first observations are briefly described, including the measurement of the Fe K-shell lines in a decaying X-class flare, hard X-ray emission from high in the solar corona, and full-disk hard X-ray images of the Sun.

\end{abstract}

\section{Introduction}
\label{sec:intro}

Understanding the origin, propagation and fate of solar high energy electrons is an important topic in solar and space physics; it is such particles that present a danger to spacecraft and astronauts in low earth orbit. These particles also carry diagnostic information that may teach us about acceleration processes elsewhere in the heliosphere and throughout the universe. Non-thermal particles can be detected via their X-ray or gamma ray emission when interacting in the chromosphere or lower solar corona, by their radio emission in the solar corona and inner heliosphere, or directly in situ in interplanetary space.

To date, the state of the art in solar hard X-ray (HXR) imaging has been the \textit{Reuven Ramaty High Energy Solar Spectroscopic Imager} \citep[\textit{RHESSI,}][]{Lin_2002}, which uses rotation modulation collimators to indirectly image the Sun. This requires detectors with large collecting areas and can provide high ($\sim$2$''$) angular resolution but is limited by both the detector background and by the nature of the Fourier imaging technique itself. In contrast, focused imaging of the Sun has been the standard in solar soft X-ray imaging for decades with detailed and dynamic images returned by the soft X-Ray Telescope (XRT) on \textit{Hinode} \citep{Golub_2007}, the Soft X-ray Telescope (SXT) on \textit{Yokhoh} \citep{Ogawara_1991}, and the Solar X-ray Imager (SXI) on the \textit{Geostationary Operational Environmental Satellites} \citep[\textit{GOES} 12-15,][]{Hill_2005}. However, these X-rays are generally emitted by lower energy thermal plasmas and the observations therefore do not directly address questions of particle acceleration \citep[see e.g.,][]{Fletcher_2011, Holman_2011}

Here we present the first focused hard X-ray images of the Sun from the \textit{Nuclear Spectroscopic Telescope ARray} \citep[\textit{NuSTAR},][]{Harrison_2013} which observes the sky from 3 to 79 keV with focusing optics whose point-spread function (PSF) has a half-power diameter (HPD) of $\sim$60$''$. \textit{NuSTAR's} primary science goals concern supernovae, black holes, and pulsars, but, unlike virtually every other high energy astrophysics mission to date, it is capable of being pointed at the Sun, observing a $12'\times12'$ patch of the solar disk at a time. Moving to focusing optics probes a completely different observing regime than has been previously possible with \textit{RHESSI}.

To compare the performance of \textit{NuSTAR} and \textit{RHESSI} we construct a figure of merit (FoM)
that contrasts the capabilities of the two different technologies by comparing the background to the effective area
of each instrument over the the energy range relevant for faint solar flares (4--15 keV). In this range the \textit{NuSTAR} average
effective area (600 cm$^{2}$ for both telescopes combined) is over ten times the \textit{RHESSI} average effective area 
(45.6 cm$^{2}$ for 8 \textit{RHESSI} detector modules). Even more importantly, the background for
the \textit{NuSTAR} detectors (8$\times 10^{-4}$ cts sec$^{-1}$) is down by over four orders magnitude compared to the \textit{RHESSI}
background \citep[54 cts sec$^{-1}$;][]{Smith:2002fx}. This results in a FoM for \textit{NuSTAR} (1.3$\times10^{-6}$ cts sec$^{-1}$ cm$^{-2}$)
roughly a million times lower than the comparable FoM for \textit{RHESSI} (1.2 cts sec$^{-1}$ cm$^{-2}$),
meaning a flare giving the same count rate as the background is a million times fainter for \textit{NuSTAR} than for \textit{RHESSI}.
While interpreting this FoM in terms of the sensitivity of the two fundamentally different types of instruments is far from trivial, this does demonstrate the power of focusing optics.

There are, however, technical challenges during solar observations that must be met to realize any increases in sensitivity. The \textit{NuSTAR} readout electronics were not specifically designed to handle the extreme count rates produced by the Sun (which can be several orders of magnitude brighter than the astrophysical sources observed by \textit{NuSTAR}) and so cannot directly observe bright solar flares as \textit{RHESSI} does. This means that \textit{NuSTAR} will complement the existing solar observatories, extending the observations in the hard X-ray band to fainter sources and opening the door to a new exploration space of hard X-ray observations of the Sun.

The remainder of this paper is organized as follows:  In \S \ref{section:analysis} we describe the mechanics behind converting the ``astrophysical" data into a useful heliophysics format and describe some of the technical challenges presented by observing the Sun with \textit{NuSTAR} with further discussion presented in the attached Appendices. We also summarize the solar observing campaign strategy and highlight the science that we are targeting with \textit{NuSTAR} solar observations.
 In \S \ref{sec:results} we describe the first year of \textit{NuSTAR} solar observations along with some early results, though we defer a detailed discussion of some of these observations to companion papers currently in production. In \S \ref{sec:discussion} we summarize our findings and present our outlook for the future hard X-ray observations of the Sun using \textit{NuSTAR}.

  \section{Data and Analysis}
\label{section:analysis}

\subsection{The \textit{NuSTAR} observatory}

\NS \ is a NASA Astrophysics Small Explorer (SMEX) satellite launched on June 13, 2012 \citep{Harrison_2013}. It has two co-aligned X-ray optics focused onto two focal planes (FPMA and FPMB) observing the sky in the energy range from 3 to 79 keV. At X-ray energies below $\sim$20 keV the field-of-view (FoV) of \textit{NuSTAR} is defined by the physical size of the detectors. This produces a FoV of roughly 12$'$x12$'$ \citep{Harrison_2013}. The point spread functions (PSF) of the optics have a full-width half maximum of 18$''$ and a half-power diameter of 60$''$ \citep{Madsen_2015}. Though \textit{NuSTAR} was designed as an astrophysics observatory, it also has the capability of directly observing the Sun without any harm to the telescope optics and only a negligible degradation of the angular resolution of the instrument. While \textit{NuSTAR} is well calibrated over the 3--79 keV bandpass \citep{Madsen_2015}, the lower energy bound can be extended to energies as low as 2.5 keV if there is sufficient flux present (See Appendix \ref{sec:low_energy}).

There are, however, several technical issues encountered when using \textit{NuSTAR} as a solar observatory. The \textit{NuSTAR} optics are based on a Wolter-I conical approximation, which is a grazing incidence, double mirror system. A properly focused photon will reflect twice off the optics before exiting. However, it is possible for a photon to reflect only once off of either the upper conical section or the lower conical section, which we refer to as a ``ghost ray". The pattern of ghost rays on the focal plane is very distinct as shown in Figure 2 of \citet{Madsen_2015} and the left panel of Figure \ref{fig:ghostrays}. On the Sun there can be several bright sources (e.g. active regions) on the solar disk but outside the FoV that can result in a complex observed pattern of overlapping ghost ray images. This is especially problematic when an extremely bright source (e.g. a bright active region or a flaring region of the Sun) is just outside of the field of view. However, as we show in Appendix \ref{sec:ghostrays}, ghost rays primarily result in an elevated background in the FoV and can be avoided by observing the Sun when no (or few) bright active regions are present on the solar disk.

Another consideration is the throughput of the \textit{NuSTAR} focal plane detector electronics. Unlike a soft X-ray or optical CCDs, the \textit{NuSTAR} detectors are not ``clocked" at a given frame rate. Instead they are photon counting devices and respond to a trigger (e.g. a photon hitting the focal plane). The readout time for each photon arriving at the detector is 2.5 ms \citep{Harrison_2013} during which time the focal plane cannot respond to a subsequently arriving photon. This can generally be described as a non-paralyzable deadtime per event of 2.5 ms \citep{Bachetti_2015} and results in a  throughput that asymptotes toward a maximum of 400 counts per second per telescope as the incident count rate increases. In practice this can mean that fainter secondary spectral components (e.g. hot plasma from above an active region or non-thermal emissions from accelerated electrons) may produce small numbers of observed counts and be hard to detect in the presence of other HXR sources (e.g. emission from the active region itself or ghost rays from sources outside of the field of view).

Fortunately, the event shaping time is short for the \textit{NuSTAR} detectors so there may not be significant pile-up effects even for the extreme count rates encountered when observing the Sun. See Appendix \ref{sec:pileup} for further discussion of potential pile-up and steps that can be taken to mitigate pile-up in the solar observations.

\subsection{Data processing}

We reduce the \textit{NuSTAR} solar data with the \textit{NuSTAR} Data Analysis Software (NuSTARDAS) version 1.4.1 and \textit{NuSTAR} CALDB version 20150414. To limit the number of solar photons that are inadvertantly vetoed in the post-processing software we disabled some of the standard filtering that is used by the NuSTARDAS pipeline. These filters are used to remove sources of electronic background noise (see the \textit{NuSTAR} Software User{'}s Guide for more information). To accomplish this we run the pipeline with the status expression {\textquotedblleft}STATUS==b0000xx00xx0xx000{\textquotedblright} and then remove obviously noisy pixels in post-processing. 

\subsection{Astrometric alignment}
\label{sec:alignment}
For the solar observations we use the ``SCIENCE\_SC" observing mode. This differs from the standard ``SCIENCE'' mode used for astrophysical sources in that it uses the aspect solution derived by the spacecraft bus (satellite) rather than the instrument star tracker to project a photon onto the sky (hereafter: ``reconstructing" the photon). This mode is automatically used when the instrument star tracker is blinded by a bright target (e.g. the bright Earth limb, the Moon, or the Sun). The spacecraft bus uses three star trackers to determine its orientation. These are pointed in roughly orthogonal directions so that at any particular moment there are combinations of one, two, or three star trackers (also know as Camera Head Units, or CHUs) that can be used to determine the aspect solution. Our absolute knowledge of the alignment of the star trackers is limited by thermal motions in the spacecraft itself. This results in a point source reconstructed using a certain combination of star trackers to appear to be shifted by an arcminute or two in RA/Dec coordinates compared to the same point source reconstructed using a different combination of star trackers. This relative shift can be removed empirically if a bright source is present in the field of view (FoV) and the two combinations can be registered against one another. The absolute RA/Dec position can be adjusted if a point source has a known position (e.g. for isolated astrophysical sources) or if the image can be registered against other observations in different wavebands (e.g. using \textit{Solar Dynamics Observatory}).

We convert the RA/Dec images to heliocentric coordinates by using the JPL Horizons online ephemeris tool \citep{Giorgini_1996} to generate the ephemeris (RA/Dec positions) for the center of the Sun and the angle between the solar north pole and celestial north. We interpolate the RA/Dec heliocentric coordinates onto the arrival times for each of the \NS \ counts and compute a differential offset (in arcseconds from the Sun center) in RA and Dec before applying a rotation to heliocentric coordinates. A final empirical offset can also be applied by aligning the \NS \ images with some other heliocentric images (e.g. AIA images from {\em Solar Dynamics Observatory}) for each CHU combination. The IDL scripts that we used to perform these operations and example solar ephemerides are publicly available via the \NS \ GitHub repository\footnote{http://www.github.com/nustar}.

\begin{table*}
	\caption{Early \NS \ Solar Observations}
	\label{tab:obs_list}
	\begin{center}
		\begin{tabular}{lccccc} \hline \hline              
			Date & Start Time &  OBSID     	&	Dwell Time  & Livetime  & Notes    \\ \hline \hline
			2014 September 10 & 21:26 & 20010001001--20010016001 & 146 sec & $<$1 sec & Offset Mosaic \\
			& 23:50 & 20011001001--20011018001 & 77 sec & $<$1 sec  & NP Slew \\
			\hline
			2014 November 1 &  16:26 & 20001002001 &  3 ks & 31 sec & North Pole \\ 
			& 18:01 & 20001003001 & 5.8 ks & 28 sec & AR 12192 Drift \\
			& 21:16 & 20012001001 & 800 sec & 3.1 sec & Dwell 1 \\
			& 21:49 & 20012002001 & 800 sec & 11.8 sec & Dwell 2 \\
			& 22:02 & 20012003001 & 800 sec & 28.5 sec & Dwell 3 \\
			& 22:15 & 20012004001 & 800 sec & 27.2 sec & Dwell 4 \\
			\hline
			2014 December 11 & 18:21 & 20001005001 & 1.6 ks & 17.1 sec & AR 12222 \\
			& 19:06 & 20001004001 & 1.5 ks & 52 sec & North Pole \\
			\hline
			2015 April 29 & 10:31 & 20110001001--20110026001 & 175 sec & $\sim$10 sec & Mosaic 1 \\
			& 12:06 & 20110030001--20110046001 & 175 sec & $\sim$10 sec & Mosaic 2 \\
			\hline \hline
			
		\end{tabular}
		
	\end{center}
\end{table*}

\subsection{\textit{NuSTAR} Solar Observation Planning}
\label{sec:science}

Observing the Sun with \textit{NuSTAR} requires detailed monitoring of the solar X-ray environment and subsequent triggering of Target of Opportunity (ToO) observations. Because of the limited field of view of \textit{NuSTAR}, the science that is accessible during each observation is directly linked to the trigger criteria for the observations, so in this section we review the science associated with a given observational trigger. A list of the \textit{NuSTAR} sequence IDs associated with the solar observations described in this work can be found in Table \ref{tab:obs_list}.

\subsubsection{Particle acceleration in solar flares}

The most direct knowledge about electron acceleration during solar flares, including the associated heating processes, is based on HXR observations \citep[e.g.,][]{Lin_2011}. Through the bremsstrahlung mechanism, HXRs provide diagnostics of hot thermal flare plasmas (typically 10-50 MK) and flare accelerated electrons (with energies above $\sim$10 keV). Solar flare HXR spectra steeply decrease with photon energy, making it challenging to observe the full spectrum. Below (typically) 10--20 keV, thermal emission from heated plasma in coronal loops dominates non-thermal emission from the footpoints and the coronal acceleration site, and generally dominates the overall count rate at all energies. \textit{RHESSI} uses entrance filters and movable attenuators to suppress the count rate from the thermal electron population, keeping the detector live time high and reducing pile-up contamination when observing the fainter high energy components. An alternative approach is to wait to observe until the solar disk itself blocks the emission from the footpoints and the thermal loops and to observe the emission from the non-thermal population of electrons that radiates at higher altitudes.

\textit{NuSTAR's} sensitivity provides the opportunity to directly observe the acceleration region for faint limb-occulted flares. HXR emission from above the flare loop could originate from the energy release of the solar flare \citep[e.g.,][]{Masuda_1994,Krucker_2010} and thus could provide new insights into the physical processes involved. While it is currently well established that the high-energy ($E>20$~keV) end of the accelerated electron population forms a power law, the low energy end is currently unexplored \citep[e.g.,][]{Krucker_2013}. One speculation is that the flare acceleration mechanism forms a Kappa distribution \citep[e.g.,][]{Oka_2013}, but the rollover from a power-law to a Maxwellian distribution at low energies ($E_{C}<15$ keV) has not been observationally confirmed. Extrapolating from \textit{RHESSI} observations of partly occulted flares \citep[e.g.,][]{Krucker_2008}, \textit{NuSTAR}, with its enhanced sensitivity, should see emission from above the main flare loop from \textit{GOES} B and C class flares, making it likely to catch events even in the decaying phase of the solar cycle.

Observing such an occulted region with \textit{NuSTAR} requires a high degree of planning and luck as the amount of occultation is crucial; if even a small fraction of the main thermal loop is visible, then \textit{NuSTAR} will be overwhelmed with counts from the soft X-ray source.  Long observations of the solar limb one to three days after a productive active region has rotated out of view are the best approach for addressing this science goal. Both the 2014 November 1 and 2014 December 11 ToO observations included targeting bright active regions just over the limb as part of the observing campaings.

\subsubsection{Heating the solar corona}
\label{sec:nanoflares}

The magnitudes of solar flares span many decades in a wide variety
of observables. Large flares \citep[and their weaker counterparts ``microflares"; see][]{Lin_1984}
tend strongly to occur in active regions, near sunspots \citep[e.g.,][]{Hannah_2011}. However, the
population of solar flares, from large flares to microflares, does not contain enough
energy to maintain the hot corona of the Sun \citep[e.g.,][]{Rosner_1978}.
\cite{Parker_1988} proposed that innumerable ``nanoflares"  \citep[a term which has now been redefined to refer to faint flares in general, e.g.,][]{Klimchuk_2006}
could produce episodic energy input that would appear as a steady heat
input to the solar corona.

Some recent studies \citep[e.g.,][]{Brosius_2014, Caspi_2015} have provided
observational evidence for nanoflare heating within active regions, 
using soft X-ray and extreme ultraviolet observations to detect hot ($\sim$5--10 MK)
plasma that is not predicted by competing heating models. However, evidence
must be weighed together with well-established, though limited, X-ray spectroscopic
observations of active regions that show no such evidence \citep[e.g.,][]{DelZanna_2014}.
The temperature structure of active regions thus remains an open question.

Several studies have suggested the existence of discrete heating
events in the quiet Sun outside of active regions \citep[e.g.,][]{Krucker_1998,Parnell_2000,Aschwanden_2000},
but evidence of associated non-thermal particle acceleration has not yet been detected.
The HXR ($E>3$ keV) is the ideal band in which to search for such a signal, though
dedicated solar HXR observatories such as
\textit{RHESSI} have lacked the necessary sensitivity to detect individual events
in the quiet sun \citep{Hannah_2010}. 

If Parker's nanoflares were to heat the corona, they would have to have 
physical properties different from microflares or flares \citep[e.g.,][]{Hudson_1991}.
Perhaps the ratio of non-thermal to thermal particle populations is smaller
in weaker flares \citep[see e.g.,][]{Hannah_2011}. Establishing this
would enable us to identify the nanoflare and flare populations
as different branches of the same physical family, with microflares
predominantly occurring in active regions while nanoflares occur
over the entire solar disk.

Searching for individual flares or nanoflare distributions in the quiet Sun requires
on-disk pointings when few active regions are present on the disk. One such
opporutnity presented itself on 2015 April 29, when the GOES full-disk X-ray flux
dipped to an extremely low level. We triggered a ToO observation that surveyed the
entire solar disk, producing the first full mosaic of the Sun in focused hard X-rays. In addition, we
dedicated several orbits during the 2014 November 1 and 2014 December 11 ToOs to targeting
``quiet" regions on the solar disk to search for a nanoflare signal.

\section{Results}
\label{sec:results}

 \begin{figure*}
\begin{center}
\includegraphics[width=0.6\columnwidth]{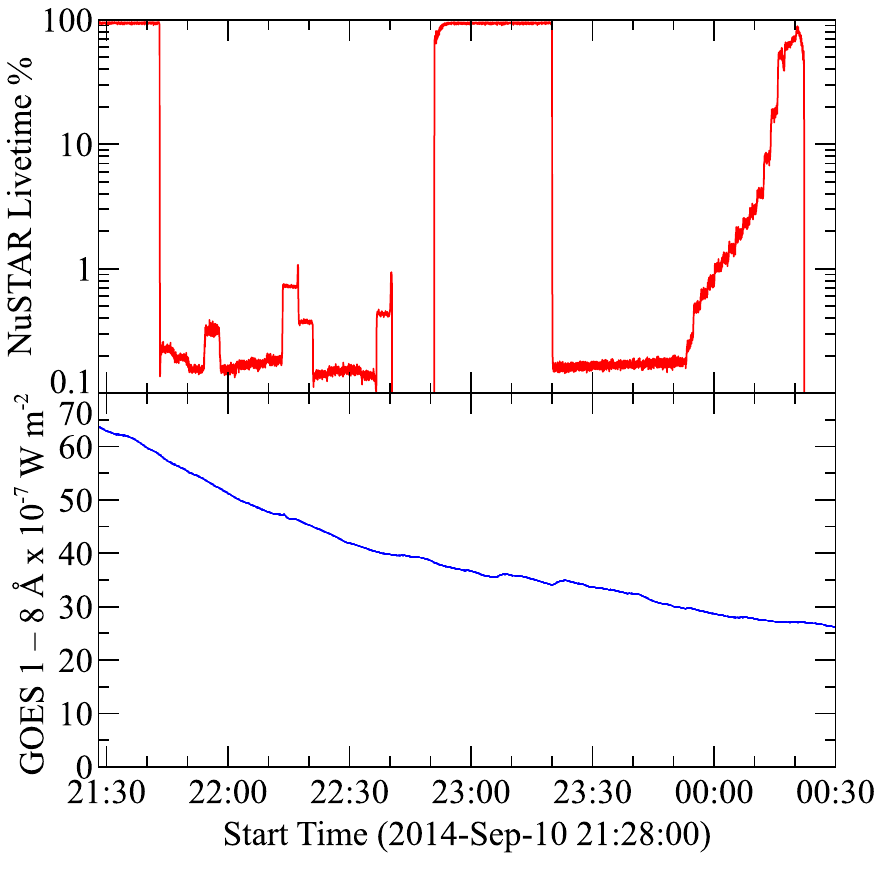}
\includegraphics[width=0.7\columnwidth]{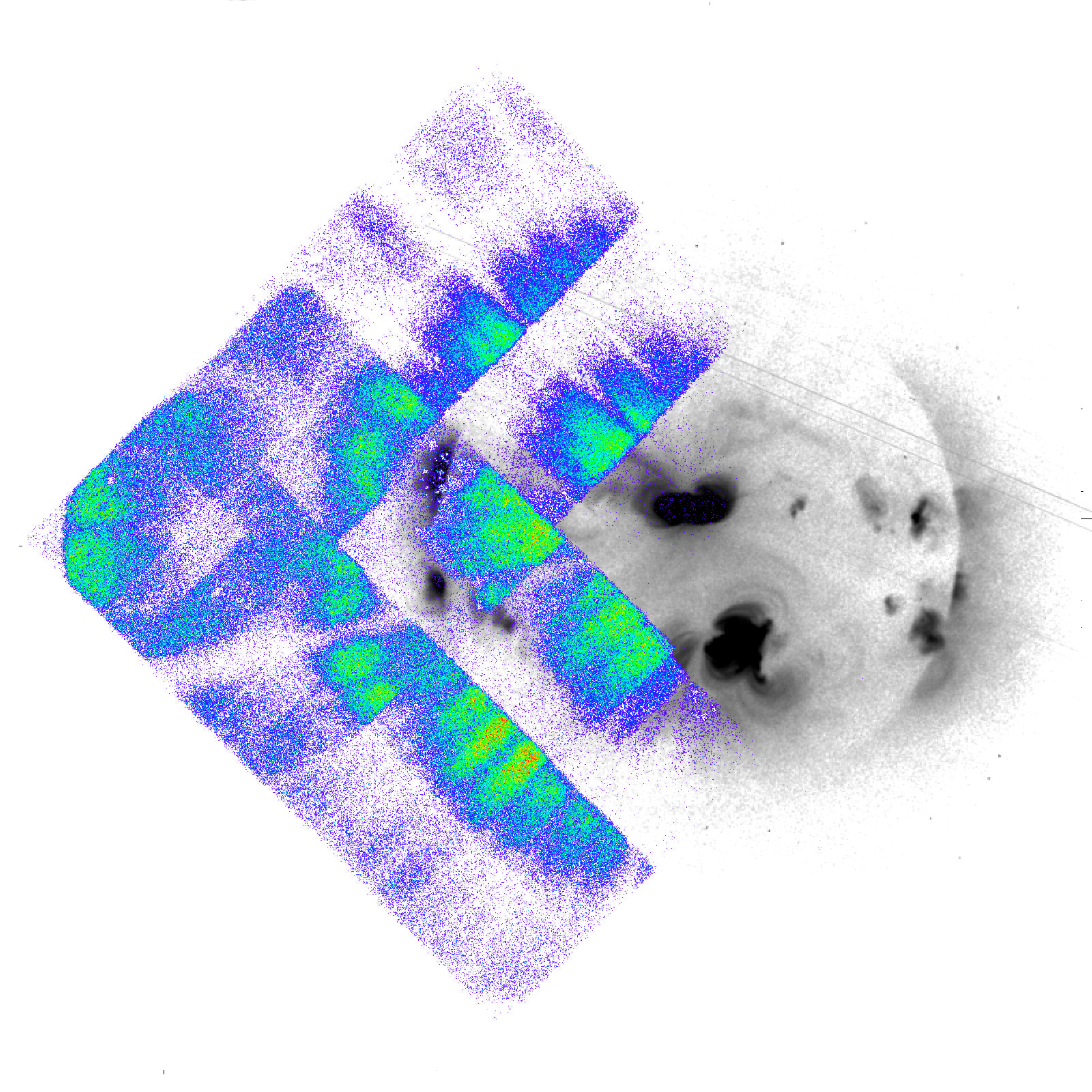}
\includegraphics[width=0.75\columnwidth]{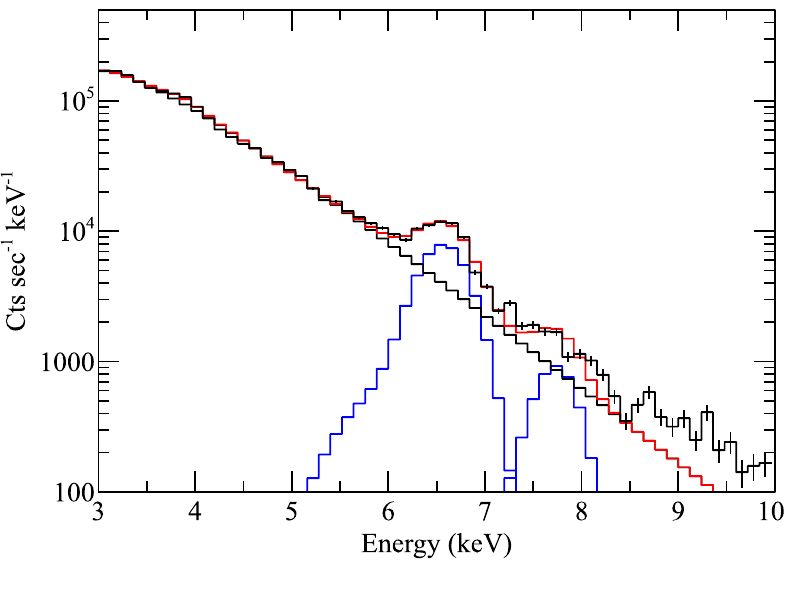}

\caption{\label{fig:obs1}
A summary of the first \NS \ solar observation. {\em Left:} The {\em NuSTAR} livetime percentage (top, red) during the decaying phase of the X-class flare along with the \textit{GOES} full-disk rate (bottom, blue). The fields of view of the two instruments are not the same, so a direct comparison between them is not straightforward. However, the \textit{GOES} count rates are useful for identifying periods of time when obvious flares occurred on the solar disk. The high livetime periods occur when the Sun is occulted by the Earth, while the \NS \ dropout near 22:45 is caused by an SAA passage. {\em Center:} The {\em NuSTAR} mosaic image (colors) showing all counts $>$2.5 keV} overlaid against the SXI Be-filter image (greyscale) showing the {\em NuSTAR} ghost rays from the flare. \textit{Right}: The {\em NuSTAR} integrated spectrum (black histogram) showing that even in this high-rate regime the {\em NuSTAR} data can be well-represented by a thermal bremsstrahlung continuum emission (black) along with emission lines from ionized Fe and Mg (blue Gaussians). The \textit{NuSTAR} data and models have been sampled at 120 eV bin widths for visual clarity.
\end{center}
\end{figure*}

\subsection{Observation 1: A pilot observation and an X-class flare}
  
A pilot observation was undertaken on 2014 September 10, and was planned to be a mosaic of the solar disk as an engineering test to ensure that no damage would come to the optics or the spacecraft by pointing at the Sun. Unfortunately, a few hours prior to the start of the observation an X-class flare erupted from the Sun. For mosaic tiles near the flaring region the event rate became so high that many photons were arriving at different spatial locations on the focal plane detectors during each 2 microsecond trigger window. This mimics the behavior of cosmic ray interactions in the detector, which are automatically vetoed by the onboard readout electronics. This resulted in the instrument being completely paralyzed during parts of the observation. Though the observation did not prove scientifically valuable, it was successful in demonstrating that even in extreme solar conditions that there was no danger to the observatory (optics, detectors, or spacecraft) due to thermal loading when pointed at the Sun.

The observation lasted for two \textit{NuSTAR} orbits. The first orbit was spent obtaining a 16-tile raster pattern with marginally overlapping fields-of-view while the second orbit was spent targeting the solar north pole and slowly slewing away from the solar disk. Both pointing strategies were designed for a much quieter solar environment and neither provided useful scientific data. An error in calculating the pointing locations led to an offset of the center of the raster pattern from the solar center which, serendipitously, allowed us to image the ghost ray pattern from the flaring regions (Figure \ref{fig:obs1}, see Appendix \ref{sec:ghostrays} for details on ghost rays). All of the photons seen in this image are ghost rays originating from the decaying X-class flare near the center of the Sun, with the apparent ``halo" patterns caused by changes in the instrument livetime and event selection as the FoV moved away from the solar disk. We find that even in this extreme case the integrated spectrum can be represented by a thin-target bremsstrahlung continuum plus the emission lines from an ionized plasma (Figure \ref{fig:obs1}, right panel).

   \begin{figure*}
  \begin{center}
\includegraphics[width=0.70\columnwidth]{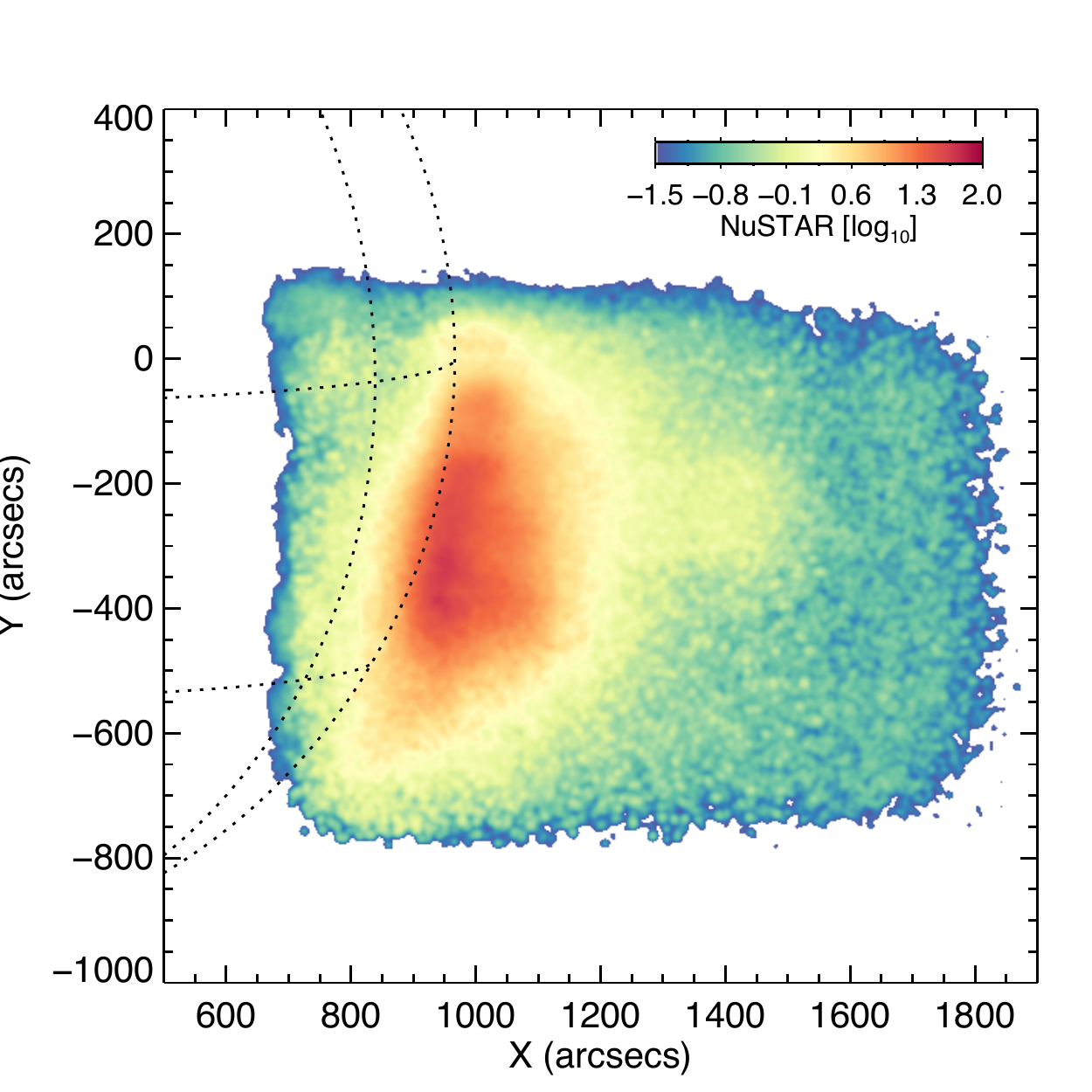}
\includegraphics[width=0.70\columnwidth]{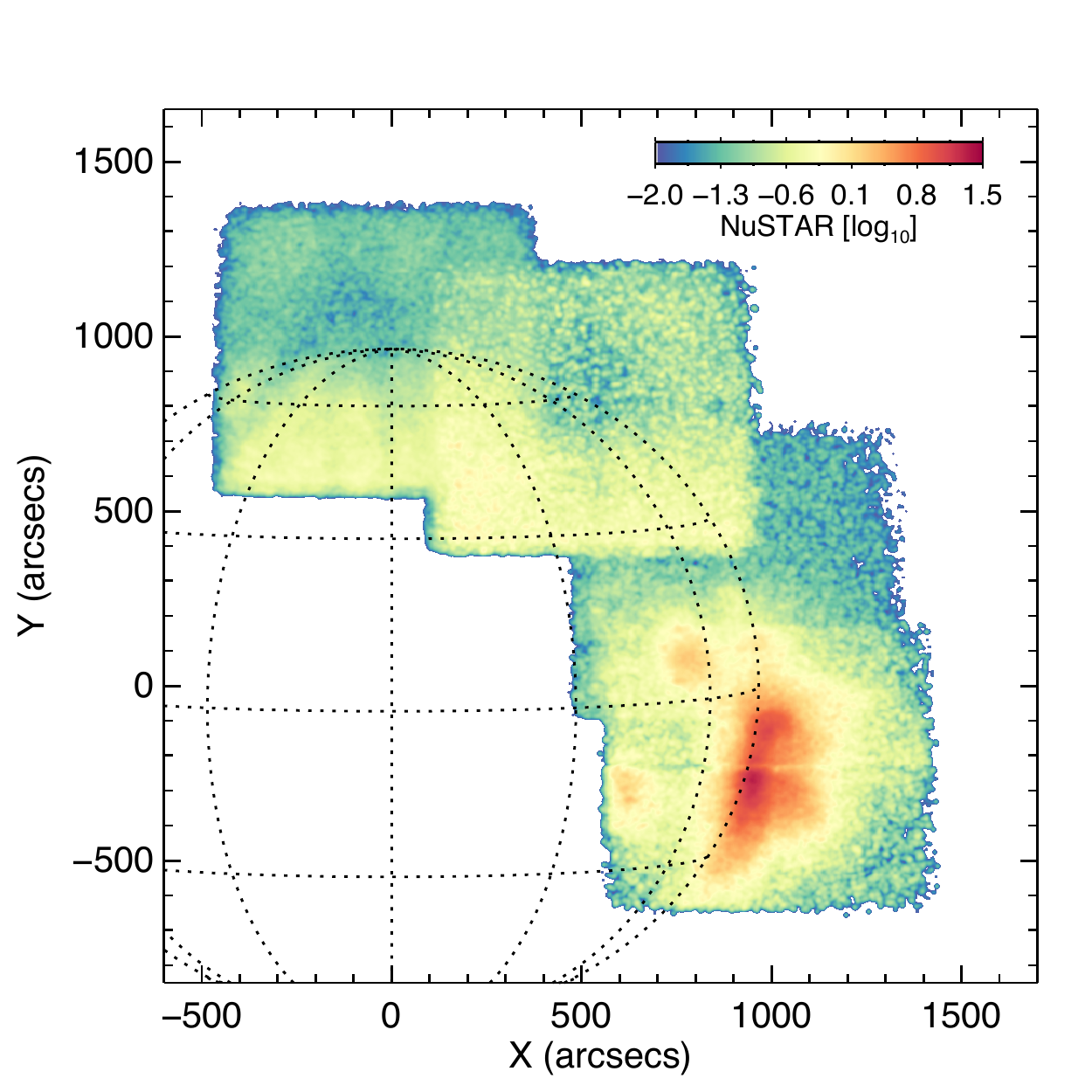}
\includegraphics[width=0.62\columnwidth]{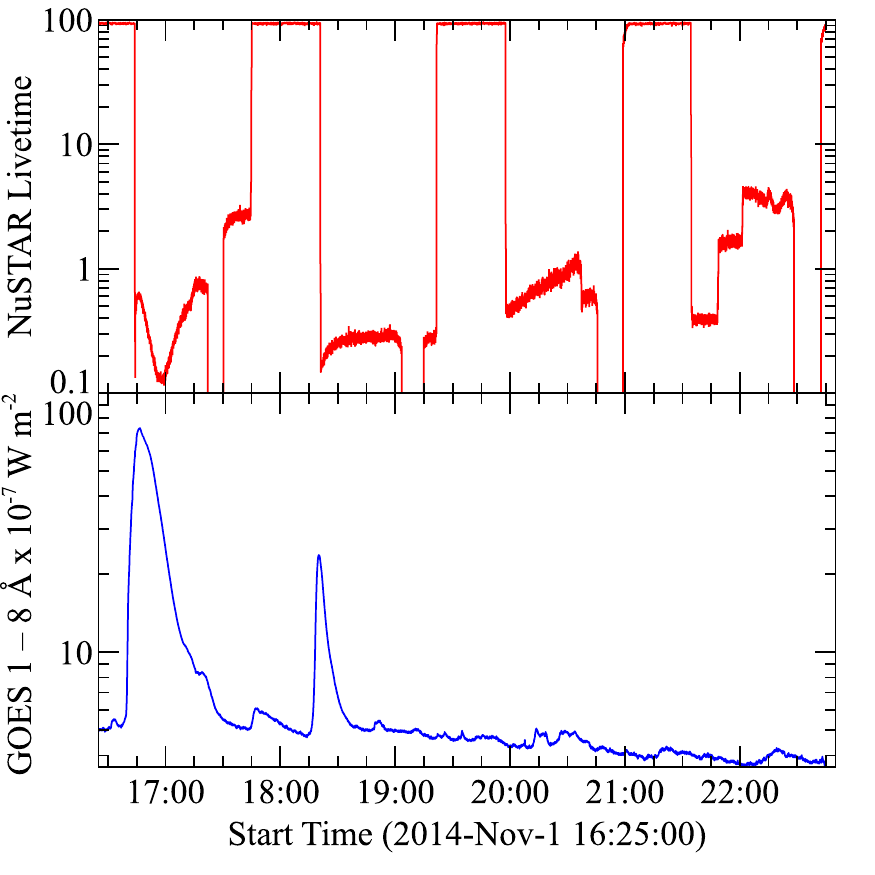}
\caption{\label{fig:obs2}
Intensity maps for the second observation showing (\textit{Left}) the \NS \ AR12192 drift orbits and (\textit{Center}) the four limb dwells. The images show all counts with energy $>2.5$ keV and the images have been smoothed by a Gaussian with a radius of 2 pixels (5$''$) to reduce statistical noise. The data have been scaled by the livetime (exposure) and are shown with a logarithmic color scale in counts per second per pixel. \textit{Right}: \textit{NuSTAR} livetime (top, red) and the \textit{GOES} full disk count rate showing the flare that washed out the North Pole pointing (around 1700 UTC and not included in the images) and the decrease in the \textit{NuSTAR} livetime in response to ghost rays from the flare.
}
\end{center}
\end{figure*}
\begin{figure*}
\begin{center}
\includegraphics[width=0.95\columnwidth]{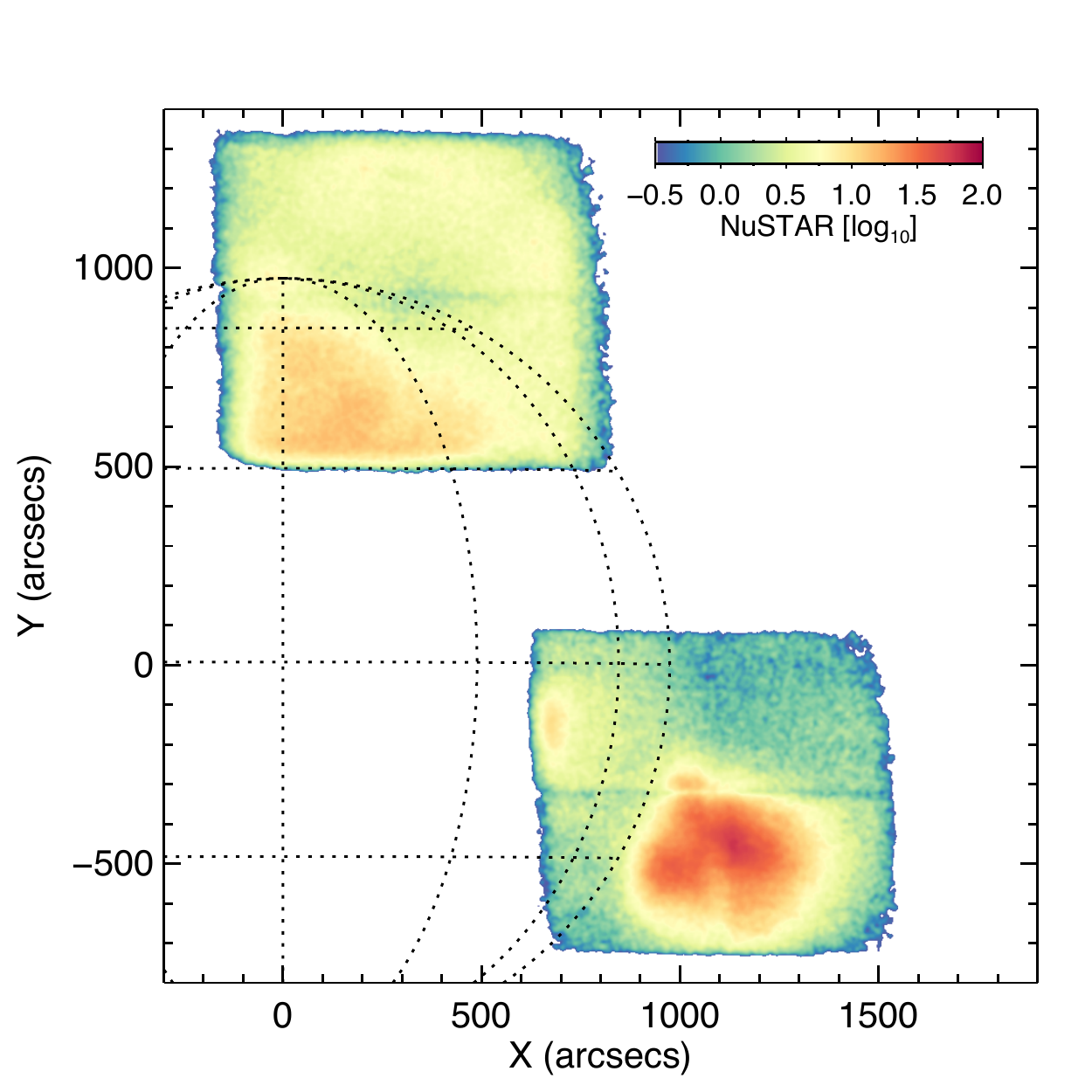}
\includegraphics[width=0.88\columnwidth]{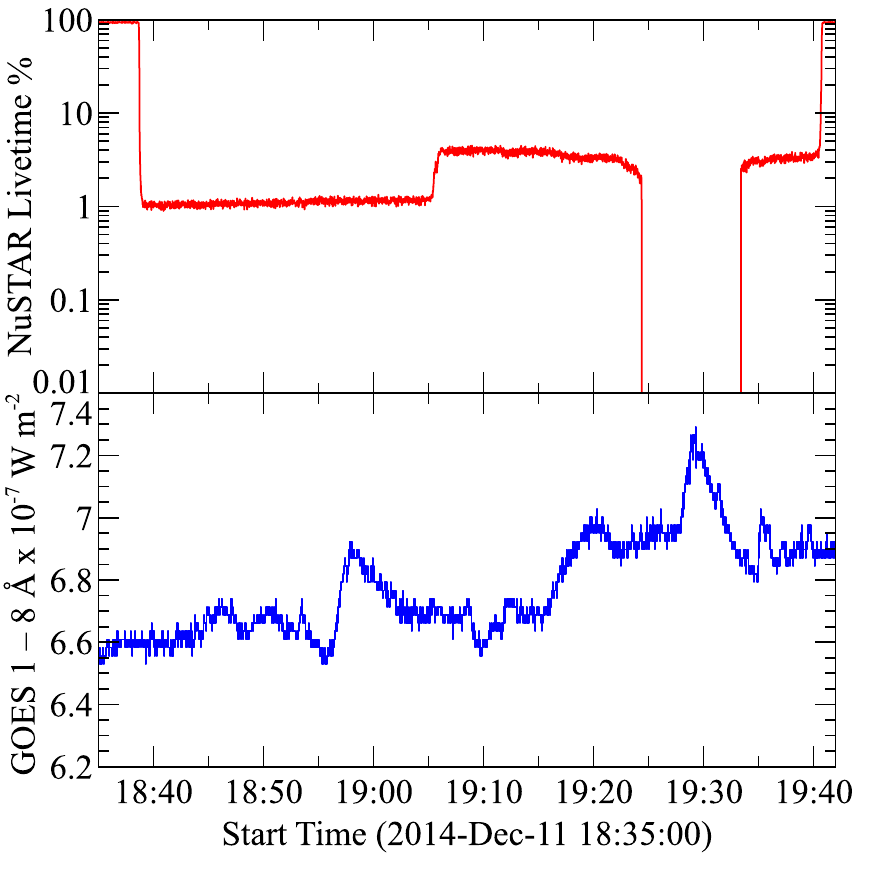}
\caption{\label{fig:obs3}
A summary of the third \textit{NuSTAR} observation on 2014 December 11. \textit{Left}:  \textit{NuSTAR} image with the data smoothed by a two pixel (5$''$) Gaussian, scaled by the instrument livetime, and shown on a logarithmic color scale. \textit{Right}:  \textit{NuSTAR} livetime percentage (top, red) along with the \textit{GOES} flux (bottom, blue). The high \textit{NuSTAR} livetime periods in the top panel are when the Sun is occulted by the Earth. The dropout in the \textit{NuSTAR} livetime near 19:30 is when the instrument is turned off as it traverses the South Atlantic Anomaly.
}
\end{center}
\end{figure*}

\subsection{Observation 2: active region 12192}

The second observation was triggered on 2014 November 1 as a bright active region (AR12192) was setting over the west limb of the Sun. The purpose was to study high coronal sources above the active region when the base of the active region were occulted by the limb of the Sun. The observation lasted for four orbits, with the first and last orbits spent observing {\textquotedblleft}quiet{\textquotedblright} regions of the solar disk. Figure \ref{fig:obs2} shows a summary of the observation, including the reconstructed \textit{NuSTAR} solar images.

As can be seen from the \textit{GOES} full-disk count rate, several microflares occurred during the first orbit, which was a dedicated pointing to the solar north pole. This resulted in this field only imaging the ghost rays from the microflares outside of the FoV (this image is not shown in Figure \ref{fig:obs2}). Fortunately the remaining three orbits resulted in a quieter solar environment.

For the middle two orbits the \textit{NuSTAR} pointing position was kept fixed with respect to the background stars, which effectively allowed the Sun to drift across the FoV. When the active region is directly in the FoV \textit{NuSTAR} predominantly samples photons from the bright thermal regions at low altitudes, while as these regions drift out of the FoV \textit{NuSTAR} can see the (relatively fainter) emission from higher altitudes. As can be seen in left panel of Figure \ref{fig:obs2}, we clearly see emission coming from over the limb of the solar disk, as well as a source of HXR at high altitudes. This source does not move with respect to the Sun as the telescope pointing drifts (which would happen if it were a ghost image from a source outside of the FoV), so we consider this to be a real source of HXR high in the solar corona that may be a remnant from flares and/or eruptions from the occulted active region. Unfortunately, the source is outside of the FoV of the \textit{GOES}-SXI simultaneous images as well as that of \textit{Solar Dynamics Observatory} (\textit{SDO}). A detailed spectroscopic analysis of the source will be address in future work.

In addition to the high coronal source, we imaged several active regions in the two southwest tiles and performed dedicated dwells targeting the north pole and the northwestern limb of the Sun to search for transient events (middle panel of Figure \ref{fig:obs2}). A spatially-resolved spectroscopic analysis of the active regions has been presented in a companion paper \citep{Hannah:2016}, while a search for transients from the quiet-Sun dwells will be presented in future work.

  \begin{figure*}
\begin{center}
\includegraphics[width=0.95\columnwidth]{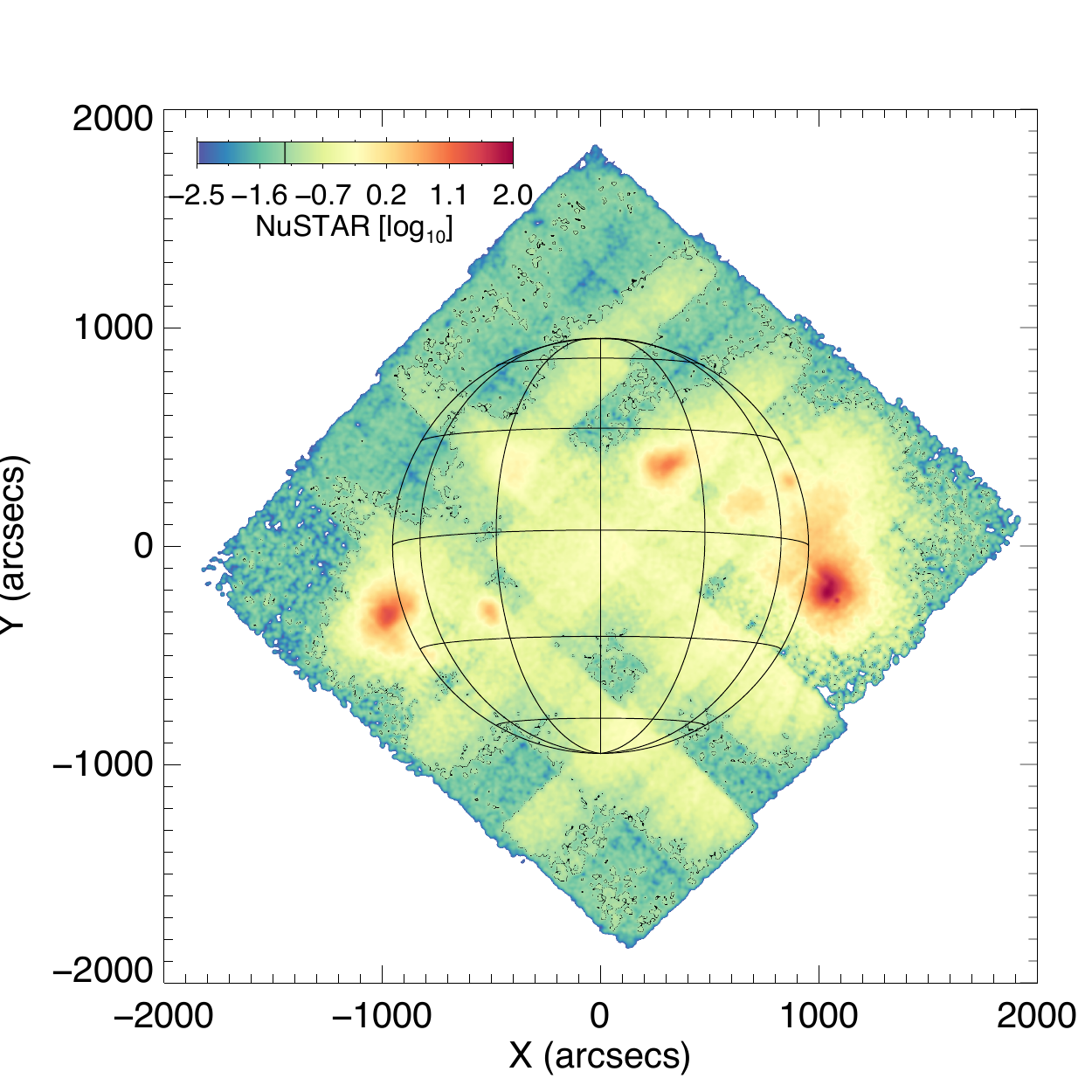}
\includegraphics[width=0.95\columnwidth]{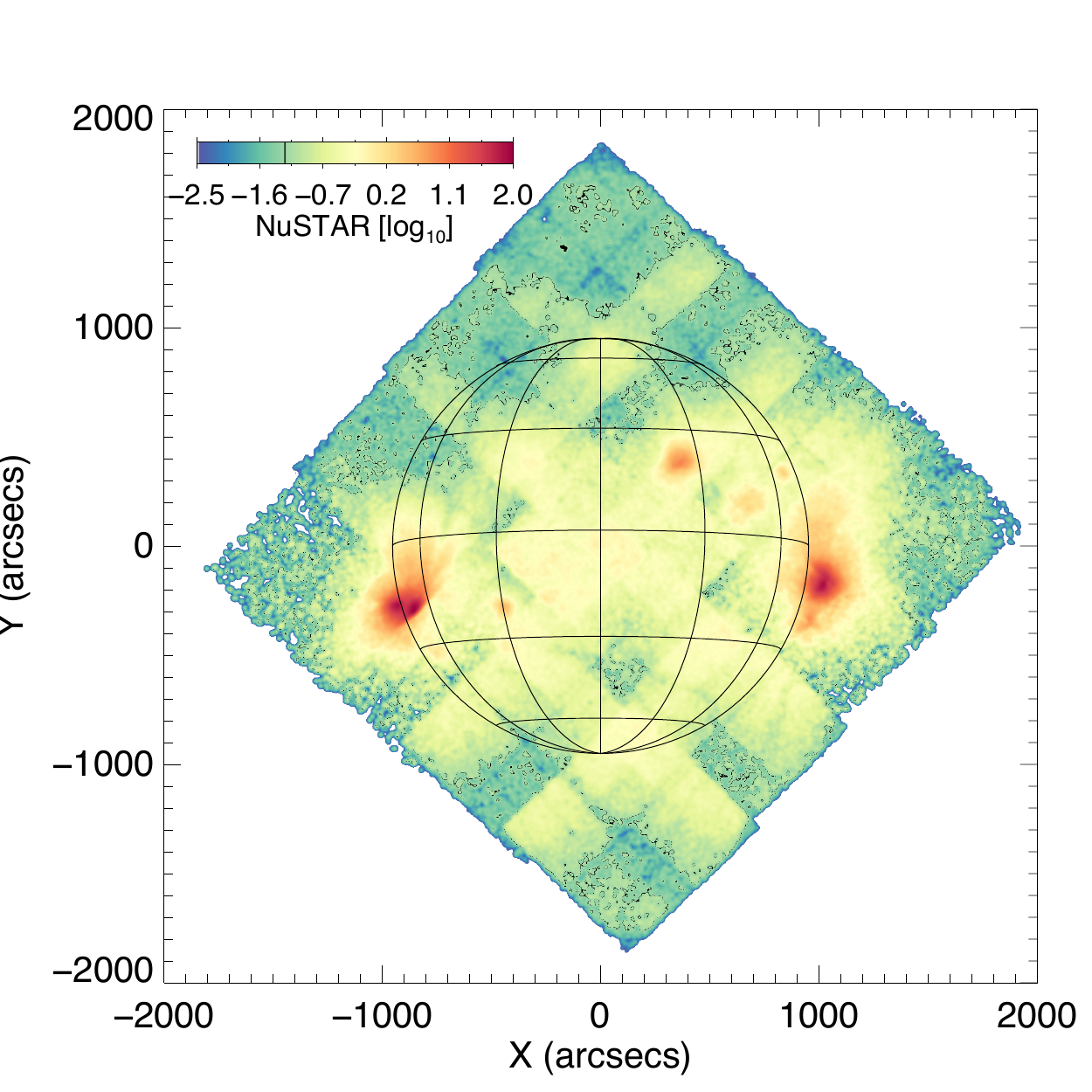}
\caption{\label{fig:obs4}
Intensity maps for the fourth \textit{NuSTAR} solar observation. The \textit{NuSTAR} images show the data for mosaic 1 (\textit{Left}, covering a time range from 10:50 to 11:50 UTC on 2015 April 29) and mosaic 2 (\textit{Right}, covering a time range from 12:27 to 13:27 UTC on 2015 April 29). The livetime corrections are applied tile-by-tile to account for variations in the count rate as the FoV moves over the solar disk. The data from both telescopes were combined and smoothed with a 2-pixel (5$''$) Gaussian and are shown on a logarithmic color scale.
}
\end{center}
\end{figure*}

\begin{figure}
\begin{center}
\includegraphics[width=0.75\columnwidth]{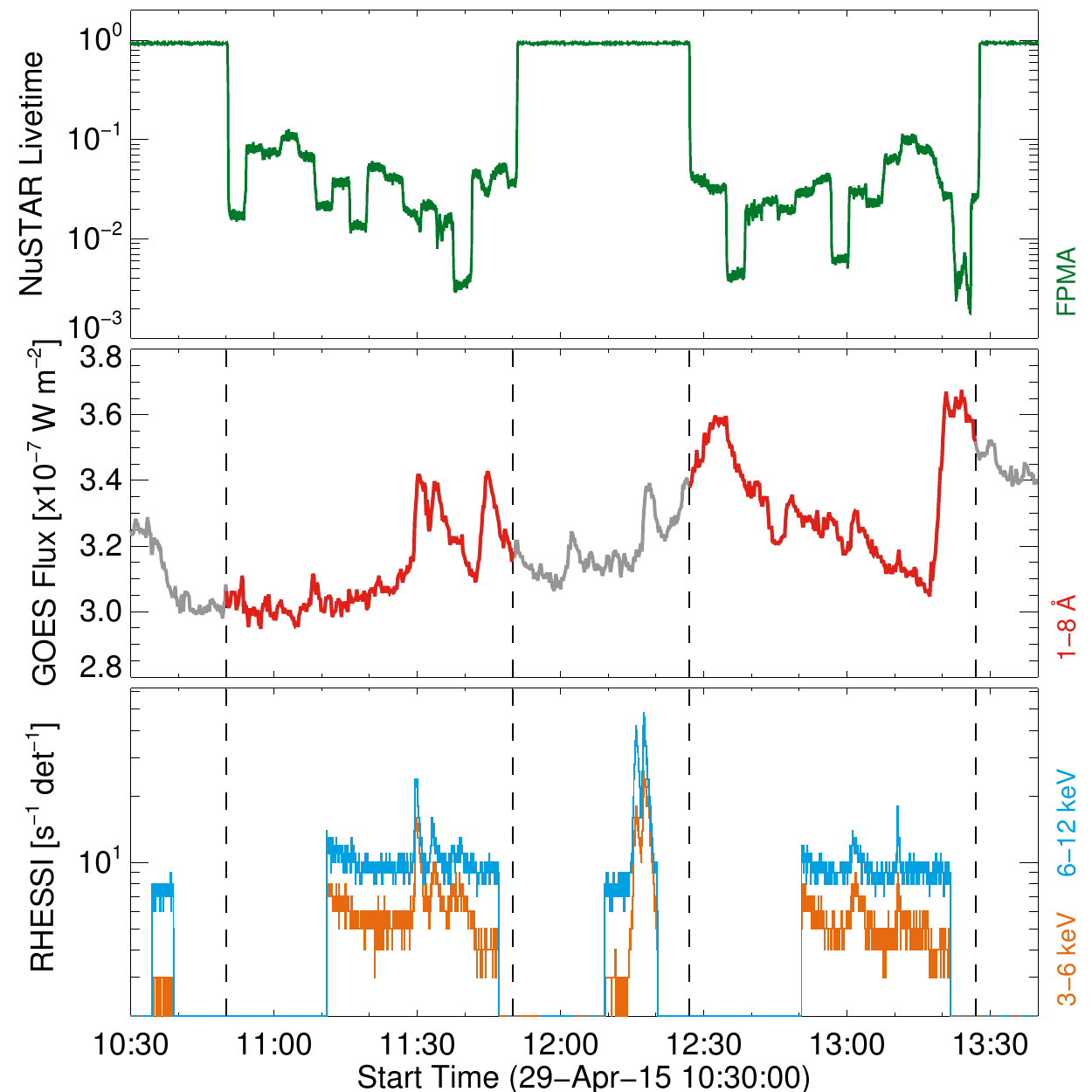}
\caption{\label{fig:obs4_livetime}
A summary of the solar mosaic observation. \textit{Top}: The \textit{NuSTAR} livetime fraction, averaged between FPMA and FPMB. \textit{Middle}: The \textit{GOES} full-disk flux. \textit{Bottom}: The \textit{RHESSI} full disk count rate. The time periods for the various microflares responsible for some of the ghost rays seen in the mosaic can be identified in the \textit{GOES} and \textit{RHESSI} data sets. The high livetime periods (near unity) for \textit{NuSTAR} occur when the Sun is occulted by the Earth (also shown as dashed lines in the bottom two panels). The full-disk monitoring capabilities of \textit{GOES} and \textit{RHESSI} are used to confirm that no flaring activity is occuring outside of the NuSTAR field-of-view that could potential produced ghost rays and contaminate the \textit{NuSTAR} data.
}
\end{center}
\end{figure}

\subsection{Observation 3: The quiet sun}

The third solar observation took advantage of a quiet solar environment on 2014 December 11 (with a full-disk \textit{GOES} level of B6). This occurred during the flight of the FOXSI-2 \citep{Christe:2016fv} sounding rocket, which also targeted an active region (AR 12222) on the limb of the solar disk. Figure \ref{fig:obs3} shows the livetime during this one-orbit observation as well as the images from the \textit{NuSTAR} limb pointing and the long stare at the north pole region. Again, we clearly see emission from the occulted active region extending high into the solar corona, though in this case we did not allow the \textit{NuSTAR} FoV to drift far enough from the limb to determine whether a high coronal source was present as for the second observation. Spectroscopic analysis of the emission from the active region is ongoing and will be reported in future work, while the quiet Sun data will be included in the search for small flares.

\subsection{Observation 4: A full-Sun mosaic}

In late April 2015 the Sun entered a quiet period with only a few relatively small active regions on the solar disk. We triggered a \textit{NuSTAR} observation to capture the first \textit{NuSTAR} full-Sun mosaics on 2015 April 29. Each mosaic lasted for one orbit and consisted of 16 tiles covering a 4 x 4 raster pattern on the Sun with each tile separated by $\sim$10$'$. A 17th tile with the FoV centered on the center of the Sun was added to each mosaic to enhance exposure at the center of the Sun to search for an axion-like signal \citep[e.g.,][]{Hannah_2010}. The  mosaic images from the raster patterns are shown in Figure \ref{fig:obs4}, while a summary of the \textit{NuSTAR} livetime and full-disk \textit{GOES} and \textit{RHESSI} count rates is shown in Figure \ref{fig:obs4_livetime}.

As in the other \textit{NuSTAR} solar observations, the images show patterns of ghost rays from sources outside of the FoV (i.e. from the active regions or flares elsewhere on the Sun). For mosaic tiles without a strong point source in the field-of-view, this results in images dominated by ghost-rays, which is responsbile for the ``checkerboard" in the combined mosaic images (e.g., Appendix \ref{sec:ghostrays} and Figure \ref*{fig:mosaic_ghostrays}).

The strongest sources within the tiles are the subject of further study, including A-level microflares that were simultaneously observed with \textit{RHESSI} and \textit{Hinode}/XRT as well as the several obvious active regions responsible for much of the ghost ray flux. There are also periods of time when the active regions outside of the FoV were not flaring and producing ghost rays and first search for an axion signal in the \textit{NuSTAR} data is underway. \\

  \section{Summary and Conclusion}
  \label{sec:discussion}

These first observations have shown that \textit{NuSTAR} will be a powerful tool for heliophysics as well as for astrophysics. The transition to focusing hard X-ray optics opens a fundamentally new parameter space for sensitive hard X-ray observations of faint features on the the Sun. Realizing this increase in sensitivity requires overcoming the technical challenges of observing the Sun with a telescope designed to search out faint objects in the distance Universe. We have identified these challenges and have shown that they can either be mitigated via data analysis techniques or avoided through opportunistic observation planning. 

The one observational challenge that we cannot overcome is the impact of ghost rays from active regions outside of the FoV. Producing firm limits on the presence of nanoflares as well as other science topics that require a quiet Sun, such as a possible axion component from the sun \citep{Hannah_2010}, require a solar disk devoid of active regions. For these cases, all we need to do is wait as we are in the declining phase of the solar cycle and solar activity is decreasing. As we reach solar minimum toward the end of this decade we expect these topics to come to fruition.

We have shown that \textit{NuSTAR} is capable of productive solar observations at the current stage of the solar cycle. Many of the questions regarding particle acceleration require flares from active regions or CMEs to be launched from the Sun and as these will become less frequent as we move toward solar minimum we will continue to observe with \textit{NuSTAR} as opportunities present themselves. Work is already under way to take advantage of the ability of \textit{NuSTAR} to provide imaging spectroscopy \citep{Hannah:2016}, to search for nanoflares, and to produce preliminary upper limits on the presence of axions in the Sun. Future coordinated observations with radio observations from VLA and and the Low-Frequency Array for Radio Astronomy (LOFAR) as well as with the currently flying \textbf{solar} telescopes (\textit{Hinode}, \textit{SDO}, \textit{RHESSI}, and \textbf{IRIS}) will enhance the rich heliophysics data provided by \textit{NuSTAR} and extend the study of the Sun across the electromagnetic spectrum.

  \section*{Acknowledgements}

This work was supported under NASA contract NNG08FD60C and made use of data from the \NS \ mission, a project led by the California Institute of Technology, managed by the Jet Propulsion Laboratory, and funded by NASA. Additional funding for this work was also provided under NASA grants NNX12AJ36G and NNX14AG07G. SK acknowledges funding from the Swiss National Science Foundation (200021-140308). AJM's participation was supported by NASA Earth and Space Science Fellowship award NNX13AM41H. Part of this work was performed under the auspices of the U.S. Department of Energy by Lawrence Livermore National Laboratory under Contract DE-AC52-07NA27344. AC was supported by NASA grants NNX15AK26G and NNX14AN84G. IGH is supported by a Royal Society University Research Fellowship.

We thank the \textit{NuSTAR} Operations, Software and Calibration teams for support with the execution and analysis of these observations. This research made use of the \NS \ Data Analysis Software (NuSTARDAS), jointly developed by the ASI Science Data Center (ASDC, Italy) and the California Institute of Technology (USA). This research made extensive use of the IDL Astronomy Library (http://idlastro.gsfc.nasa.gov/) and of the SolarSoft IDL distribution (SSW). Additional figures were produced using the Veusz plotting package (\copyright~2003-2015 Jeremy Sanders).
 \\

{\em Facilities}: \textit{NuSTAR}

\appendix

\section{Low-energy spectral fitting}
\label{sec:low_energy}

\begin{figure*}[ht]
\begin{center}
\includegraphics{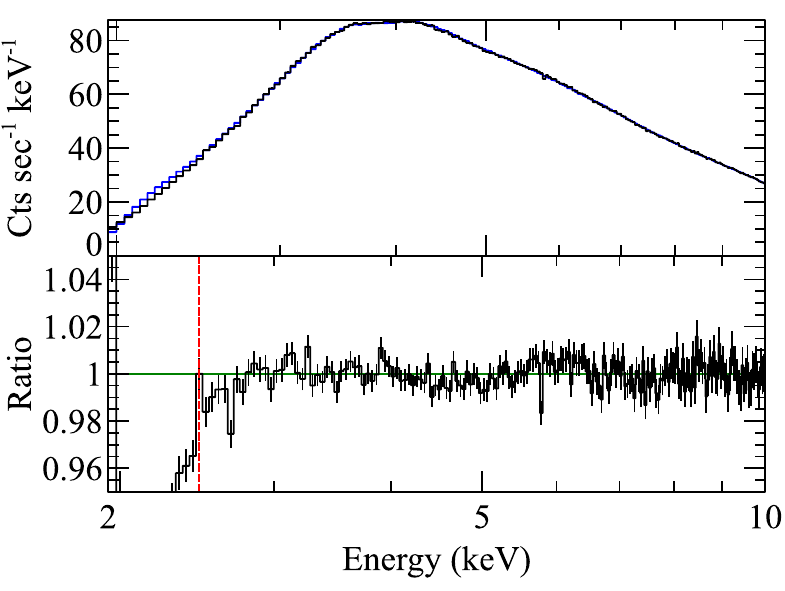}
\caption{\label{fig:crabspec}
{\em Top:} The spectrum from the Crab Nebula and Pulsar along with the fiducial model with a power law index of 2.1 and a flux normalization 8.5 ph keV$^{-1}$ cm$^{-2}$ sec$^{-1}$ at 1 keV. Interstellar absorption has corrected for with the solar abundances given by \citet{Wilms_2000} and the photoionization cross-sections given by \citet{Verner_1996}. The black histogram shows the data (error bars are present by not visible), while the blue line shows the fiducial model folded through the \textit{NuSTAR} instrument response functions. \textbf{{\em Bottom}: The black histogram shows ratio of the data to the fiducial model spectrum along with 1-$\sigma$ statistical errors.} The horizontal (green) line shows a ratio of unity, while the vertical dashed (red) line shows the 2.5 keV limit where the errors in the instrument response start to become larger than the systematic uncertainties above 3 keV (which are at the $\sim$1\% level).
}
\end{center}
\end{figure*}

The \textit{NuSTAR} effective area has been calibrated against the spectrum from the Crab Nebula and Pulsar \citep{Madsen_2015} for energies above 3 keV.  Below $\sim$5 keV the thermal blanketing surrounding the optics and the Be window that protects the detectors start to photoelectrically absorb photons, resulting in a steep decline in the throughput. For most astrophysical cases the source spectrum is faint enough that it is not practical to continue analyzing data below 3 keV. For the Sun, however, the thermal spectrum rising steeply toward low energies can produce an observed count spectrum which can frequently peak below 3 keV. At these energies uncertainties in the attenuation along the optical path start to affect the observed spectrum. If we simply extrapolate the response below 3 keV down to 2.5 keV (Figure \ref{fig:crabspec}) the systematic residuals of the data from the fiducial Crab spectrum are at the 1-2\% level, which is comparable to the overall systematic uncertainty in the instrument response below 10 keV. Below 2.5 keV there are additional systematic variations (as can also be seen in Figure \ref{fig:crabspec}) casued by variations in the trigger threshold for individual \textit{NuSTAR} pixels not captured in the instrument response files. The nominal energy threshold varies pixel-to-pixel with a mean value of $\sim$2 keV and a 1-sigma spread of 0.4 keV and so the pixel-to-pixel variations could potentially introduce artifacts in the observed spectrum below $\sim$2.5 keV. We therefore recommend limiting all spectral fitting to energies above 2.5 keV.

\section{Ghost rays}
\label{sec:ghostrays}

\begin{figure*}
\begin{center}
\includegraphics[width=0.45\columnwidth]{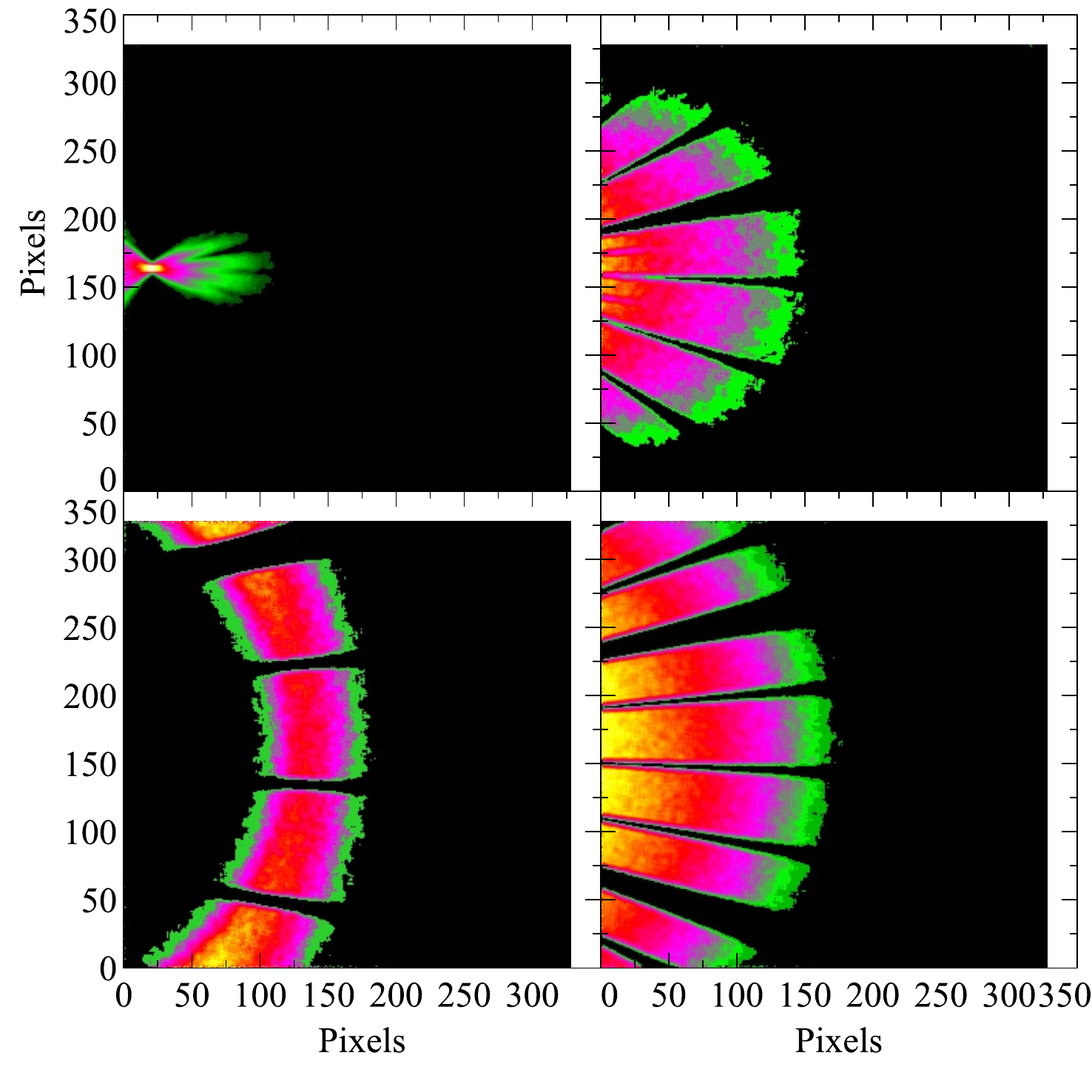}
\includegraphics[width=0.45\columnwidth]{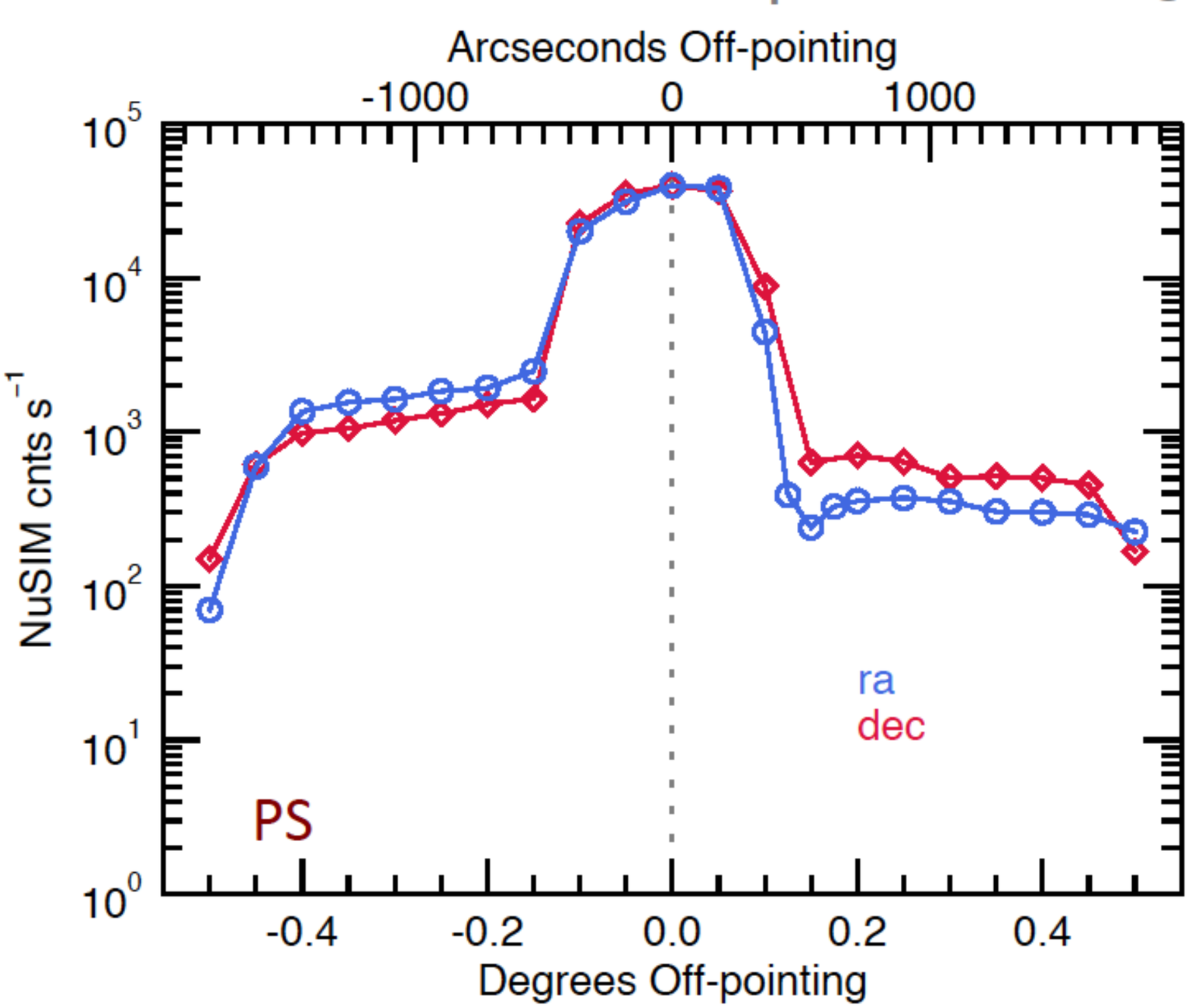}
\caption{
\label{fig:ghostrays}
{\em Left:} The simulated ghost-ray pattern from a source at different distances from the optical axis. The color scale shows the intensity, which has been allowed to vary between panels. Clockwise from the top left the panels show a point source at 6, 12, 20, and 30$'$ off-axis. The x- and y-axes are shown in simulated focal plane pixels of 122 $\mu$m, which oversamples the physical pixel size by a factor of a few but illustrates the scale of the structure in the ghost-ray pattern. {\em Right:} Contribution to the count rate when integrated over the entire focal plane as a function of off-axis angle. The data points with rates $>10^{4}$ cts sec$^{-1}$ are in the FoV (and the count rate includes the focused X-rays), while the lower count rates are outside of the FoV and are only seen via their ghost-ray pattern. The active region contributes a significant number of counts when integrated over the field-of-view even when the active region is 0.5 degrees away from the pointing location.
}
\end{center}
\end{figure*}

According to ghost ray models, a source outside of the FoV will produce ghost rays at a rate down by several orders of magnitude compared what the same source would produce on-axis (i.e. when we are observing the double-bounce ``focused" photons). Figure \ref{fig:ghostrays} demonstrates this by showing the count rate (integrated over the whole focal plane) of a ray-trace simulation that placed a bright source at a range of off-axis angles. While the source is in the FoV it produces count rates on the order of 10$^{4}$ counts per second, while the source only produces count rates on the order of 10$^{2}$--10$^{3}$ counts per second when outside of the FoV. This elevated background has spatial structure that must be accounted for during analysis. Unfortunately, the exact ghost ray image that we observe is difficult to model, since we do not directly know the distribution of sources responsible for the ghost rays, and so it's generally not possible to simulate exact ghost ray images that can be subtracted from the patterns that we observe.  However, if we know the location of the active regions (e.g. using a soft X-ray telescope to track flaring regions), then we can produce masks to screen out the regions of the detector most affected by ghost rays using the \textit{NuSTAR} ray-trace code.

\begin{figure*}
	\begin{center}
		\includegraphics[width=1.0\columnwidth]{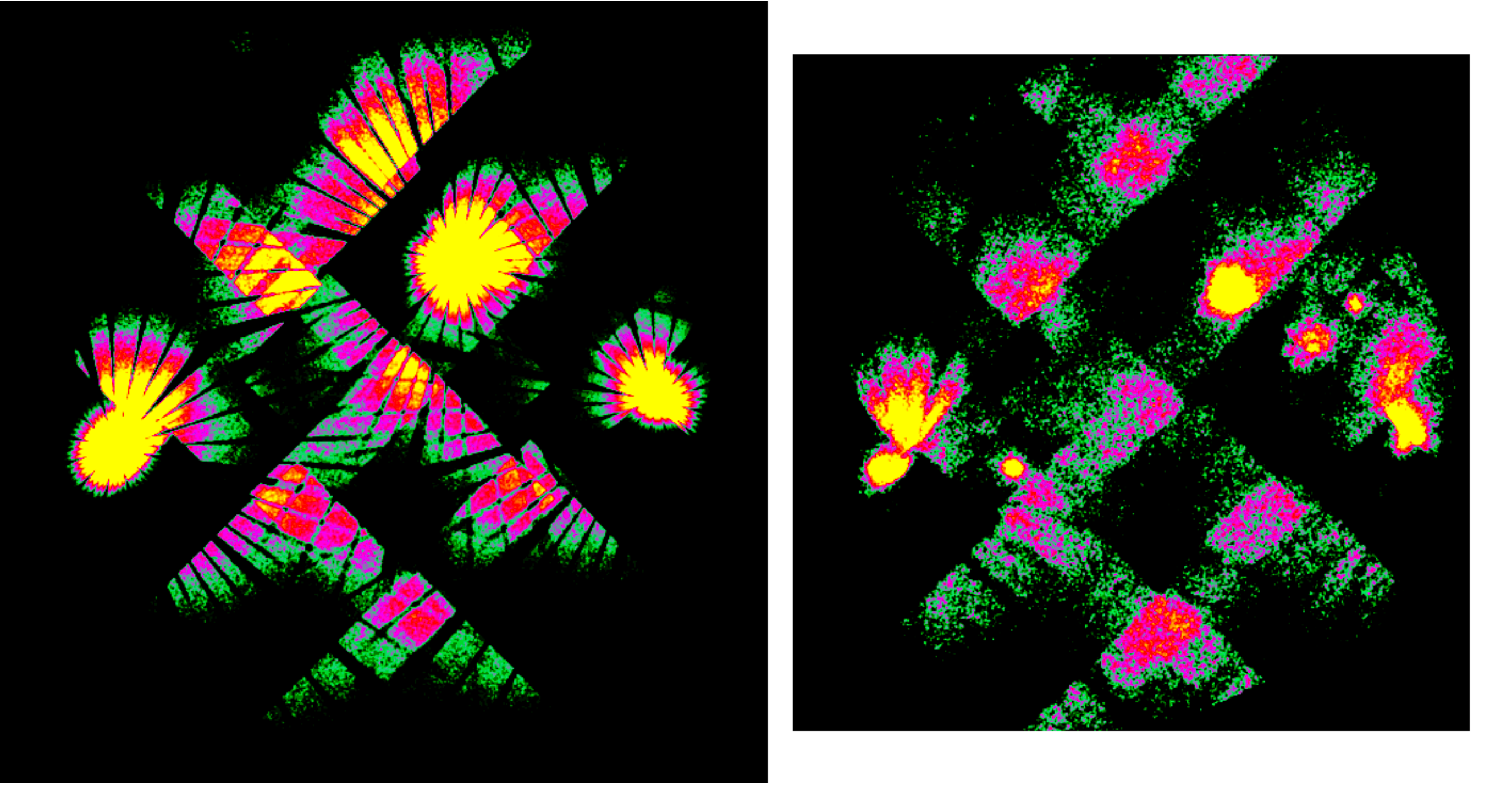}
		\caption{
			\label{fig:mosaic_ghostrays}
			A comparison of the simulated ghost-ray pattern from observed during the NuSTAR mosaic (\textit{Left}) based on the bright sources detected during the mosaic (\textit{Right}). The color-scale here is manipulated to emphasize the ghost-ray patterns between the two images but is set to have roughly the same dynamic range between the simulation and the data.		
		}
	\end{center}
\end{figure*}

For example, we can simulate the ghost-ray images observed in the \textit{NuSTAR} mosaic images above (Figure \ref{fig:obs4}). Based on the bright sources observed during the mosaic we can use our knowledge of the NuSTAR pointing location to produce simulated images of the point sources and their ghost rays (Figure \ref{fig:mosaic_ghostrays}). Unfortunately, precisely matching the ghost-ray pattern and the observed data requires knowing the instantaneous flux from each of the point sources; since by definition they're outside of the field-of-view this is not possible based on the \textit{NuSTAR} data alone and, since both \textit{GOES} and \textit{RHESSI} typically provide full-disk intensities, it's generally not possible to know the time-variable flux from flaring AR outside of the FoV.

We note that, in addition to the point sources considered here, the solar disk itself may also contribute ghost-rays in the hard X-rays. Because we do not have a good understanding of the hard X-ray flux from the solar disk, which is known only as upper limits, we must wait for quiet Sun observations that are not contaminated by ghost-rays from any point sources (e.g. active regions) to determine the effects of the solar disk. However, initial ray trace simulations indicate that the ghost ray contribution from the solar disk will be at the 10--15\% level of the flux from the solar disk itself. Such analyses will be critical in the search for emission from the solar disk and for solar axions.

\section{Event pile-up}
\label{sec:pileup}

The \NS \ detectors are relatively immune to pile-up in most astrophysics sources. These sources typically produce incident count rates of less than 1 count per second, though bright X-ray binaries can produce several hundred to several thousand counts per second on the detectors. In contrast, the observations of the Sun typically produce incident rates of several hundred thousand counts per second. In this extreme count rate case we may need to account for the effects of pile-up. The general readout scheme of the electronics is described by \cite{bhalerao2012neutron}, while the effects of deadtime on time-series analysis have been also been discussed in the literature \citep{Bachetti_2015}. Here we discuss the potential effects of any pile-up on the observed spectrum.

Pile-up can occur in \NS \ in two ways: (1) Two photons occur in the same pixel and are read out as a single-pixel event by the on-board electronics; (2) Two photons occur in adjacent pixels and are identified as a ``split-pixel'' event (Grade $> 0$) in the post-processing software and the pulse heights are combined in the post-processing software.

We investigate the first type of pile-up using the brightest astrophysical source, Sco X-1. Sco X-1 was the second source of X-rays discovered in the sky \citep[the first beyond the Sun;][]{Giacconi_1962} and has been extensively studied by nearly all X-ray observatories. It has a well-known tendency to enter a flaring state, which produces incident count rates $>10^{3}$ counts per second in \textit{NuSTAR}.  \textit{NuSTAR} observed Sco X-1 for 20 ks (sequence ID 30001040002) and detected two periods where the source entered a flaring state when the incident count rate exceeded 15,000 counts per second on the focal plane for extended periods of time as well as a period when the incident count rate was a more modest 6,000 counts per second. For reference, the Crab produces an incident count rate of roughly 1,000 counts per second. Since Sco X-1 also has a steep thermal-type spectrum we can use it as a proxy to study the effects of pile-up that we might anticipate for observations of the Sun, noting that Sco X-1's effective temperature (2--3 keV) is much higher than that of the quiet Sun, which may not exceed 0.15 keV \citep{Sylwester_2012}.

\begin{figure*}
\begin{center}
\includegraphics[width=0.55\columnwidth]{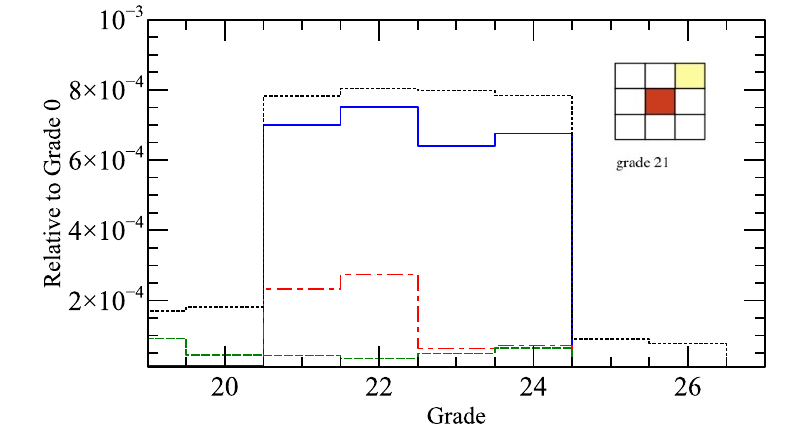}
\includegraphics[width=0.4\columnwidth]{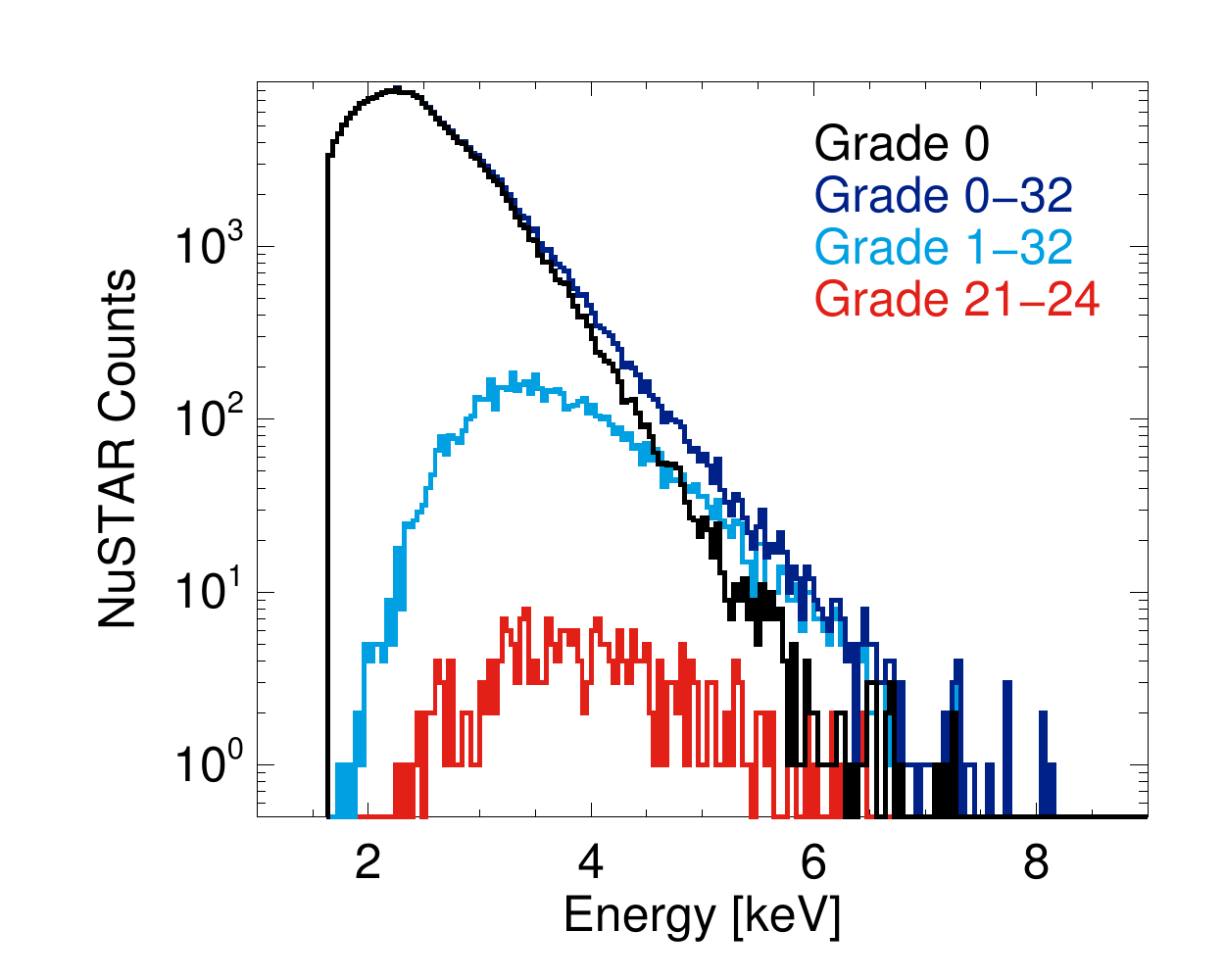}
\caption{\label{fig:gradespec}
{\em Left:} The comparison of the non-physical grade distribution for the Crab (dashed green line), Sco X-1 (dotted black line), the data when a bright active region was in the FoV (solid blue) and when only the ``quiet" Sun was in the FoV (dash-dot red line). All of the data are given relative to the number of Grade 0 events, so for the solid blue curve we expect seven Grade 21 counts for every 10,000 Grade 0 counts. An example of the non-physical grades is given in the inset, with both the central pixel and the top right pixel triggering. Grades 22 through 24 are similar with the central pixel and the bottom right, bottom left, and top left pixels triggering, respectively. {\em Right:} The spectroscopic effect of rejecting piled-up photons. 
}
\end{center}
\end{figure*}

We can investigate this by considering event ``grades" that are non-physical. A ``grade" is a qualifier assigned to each event by the post-processing software by determining which pixels in a 3x3 grid centered on the triggered pixel have collected charge above a set software threshold. ``Grade = 0" events are those where only the central pixel is above threshold (e.g. a single-pixel event), while Grade$>$0 events can occur when charge is ``shared" between pixels in the detector (for more details see the \textit{NuSTAR} User's Guide at the HEASARC). There are certain grades that cannot be produced by charge sharing (e.g. the inset in the left panel of Figure \ref{fig:gradespec}). Photons with these grades can only occur when a second photon arrives within 2 microseconds of the first photon.

We can perform a simple test to see if the number of non-physical grades that we observe for Sco X-1 matches the expectations for a given incident rate, which is described by Equation \ref{eqn:pileup}:

\begin{equation}
Prob(t > \tau;R, EEF) = 1 - e^{-\tau \cdot R \cdot EEF / 9}
\label{eqn:pileup}
\end{equation}

Here, $R$ is the incident rate [cts sec$^{-1}$], $\tau$ is the pile-up window [sec], and $EEF$ is the probability that a second photon arrives within the 3x3 detector pixel grid centered on the first photon. To estimate $EEF$ we have used the encircled energy function (EEF) files in the \NS \ CALDB to determine the probability that a second photon will arrive within a radius of ~2 detector pixels ($\sim$25 arcsecond) from the center of the PSF and divided by 9 to account for the number of pixels that could potentially cause pile-up. For $\tau$ = 8$\times10^{-6}$ seconds, $EEF$ = 5\%, then for an incident rate of 12$\times10^{3}$ counts per second we expect a pile-up fraction of 5.3$\times10^{-4}$ per pixel. If we use Grades 21-24 (e.g., inset of Figure \ref{fig:gradespec}) in Sco X-1 as a proxy we instead find a pile-up fraction of 8e-4 per pixel, which we can recover if $EEF$ is 7.5\% instead of 5\%, which is a reasonable correction as the EEF is integrated radially while in reality we are using a 3x3 square detector grid. We consider this to be a proof of concept that we have a good understanding of the origin of the pile-up in the \textit{NuSTAR} detectors and we can then use the Grade 21-24 events as a proxy to measure the pile-up fraction.

The major difference between Sco X-1 and the Sun is that Sco X-1 is a point source and so the counts will be distributed on the focal plane in a pattern described by the PSF of the optics, while for the Sun the source or sources of HXRs may be extended and may spread over the entire field of view. This implies that, except for the case where a source of solar HXRs is small enough to emulate a point source, we generally expect fewer than 0.05\% of events to be piled-up and so we can neglect the contribution of piled-up photons of the first type. However, we do urge caution and recommend that observers use this simple test to calculate the observed pile-up fraction by calculating the number of events per grade over the Grade 21--24 range.

This leaves only pile-up of the second variety where two photons arrive in adjacent pixels and are combined during post-processing. This is a far more simple case that we can simply reject in post-processing by ignoring all events with Grades$>$0. The right panel of Figure \ref{fig:gradespec} shows the effect of applying such a filter to the \textit{NuSTAR} data. The difference in slope above $\sim$5~keV between the Grade 0 (single-pixel, black) events and the ``multiple-pixel" events (cyan and red) indicates that the multiple-pixel events are in fact piled-up counts masquerading as multiple-pixel events. We view this as confirmation that we can remove the piled-up photons of the second type.

\end{document}